\documentclass[a4paper,12pt]{article} 
\usepackage[left=2.5cm,right=2.5cm,top=2.5cm,bottom=2.5cm,bindingoffset=0cm]{geometry}
\usepackage[T2A]{fontenc}
\usepackage[cp1251]{inputenc} 
\usepackage[russian,english]{babel} 
\usepackage{indentfirst}
\usepackage{amssymb,amsmath}
\usepackage{pst-all}
\usepackage{graphicx}
\usepackage[hyphens]{url} 
\usepackage{algpseudocode} 
\usepackage[off]{auto-pst-pdf}
\usepackage{hyperref}
\usepackage{amsthm}

\usepackage{color,ifpdf,graphicx}
\ifpdf 
  \DeclareGraphicsExtensions{.pdf,.png,.mps}
\else 
  \DeclareGraphicsExtensions{.eps,.bmp}
  \usepackage{pstricks}
\fi
\usepackage{pgf,epic,bez123}

\title{Switching strategy based on homotopy continuation for non-regular affine systems with application in induction motor control}

\date{}
\author{Alex Borisevich\footnotemark[1] \and Gernot Schullerus\footnotemark[2]}

\newcommand{\Lie}{{\mathcal{L}}}
\newcommand{\R}{\mathbb{R}}
\DeclareMathOperator{\const}{const}
\DeclareMathOperator{\rank}{rank}
\DeclareMathOperator{\sign}{sign}

\newtheorem{Def}{Definition}
\newtheorem{Thm}{Theorem}
\newtheorem{Lmm}{Lemma}
\newtheorem{Asm}{Assumption}
\newenvironment{Prf}[1][Proof]{\begin{trivlist}
\item[\hskip \labelsep {\bfseries #1}]}{\end{trivlist}}

\begin{document}

\maketitle

\footnotetext[1]{St. Petersburg State Polytechnical University, Polytekhnicheskaya 29, St. Petersburg, 195251, Russia, \href{mailto:alex.borysevych@spbstu.ru}{alex.borysevych@spbstu.ru}}
\footnotetext[2]{Reutlingen University, Alteburgstr. 150, 72762 Reutlingen, Germany, \href{mailto:gernot.schullerus@reutlingen-university.de}{gernot.schullerus@reutlingen-university.de}}

\begin{abstract}
In the article the problem of output setpoint tracking for affine non-linear system is considered. Presented approach combines state feedback linearization and homotopy numerical continuation in subspaces of phase space where feedback linearization fails. The method of numerical parameter continuation for solving systems of nonlinear equations is generalized to control affine non-linear dynamical systems. Application of proposed method demonstrated on the speed and rotor magnetic flux control in the three-phase asynchronous motor.
\end{abstract}

\section{Introduction}

Let the affine nonlinear system with $m$ inputs and $m$ outputs in state space of dimension $n$ is given:

\begin{equation}\label{plant}
\dot{x} = f(x) + \sum_{i=1}^m g_i(x) u_i, \; y = h(x),
\end{equation}

where $x \in X \subseteq \R^n$, $y \in Y \subseteq \R^m$, $u \in U \subseteq \R^m$, maps $f : \R^n \to \R^n$, $g_i : \R^n \to \R^n$, $h : \R^n \to \R^m$ are smooth vector fields $f,g,h \in C^{\infty}$. Functions $f(x)$ and $g(x)$ are considered as bounded on $X$.

Systems of the form \eqref{plant} are the most studied objects in the nonlinear control theory. 

There are several most famous control methods for systems of type \eqref{plant} : feedback linearization \cite{1,2,3}, application of differential smoothness \cite{4}, Lyapunov functions and its generalizations \cite{5}, including a backstepping \cite{6}, also sliding control \cite{7} and approximation of smooth dynamic systems by hybrid (switching) systems and hybrid control \cite{8}.

All of these control techniques have different strengths and weaknesses, their development is currently an active area of research, and the applicability and practical implementation has been repeatedly confirmed in laboratory tests and in commercial hardware.

Approach described below is based on the method of numerical parameter continuation for solving systems of nonlinear equations \cite{9}, which deals with parametrized combination of the original problem, and some very simple one with a known solution. The immediate motivation for the use of parameter continuation method in control problems is a series of papers \cite{10,11}, in which described the application of these methods directly in the process of physical experiments.

In this paper we consider the solution of the output zeroing problem for the system \eqref{plant} with relative degrees $r_j \ge 1$ that expands earlier obtained in \cite{12} and \cite{13} results for a case $r_j = 1$. Further it is supposed that \eqref{plant} it is free from zero-dynamics, i.e. $n = \sum_{j=1}^m r_j$.

The article consists of several parts. We briefly review the necessary facts about the method of parameter continuation and feedback linearization. Next, we represent the main result, an illustrative example of the method, as well as an example of controlling three-phase induction motor.

\section{Problem statement and motivation}

In this paper we consider the problem of nonlinear output regulation for affine nonlinear system. In particular, we will solve the problem of output regulation to constant setpoint (without loss of generality, regulation to 0).

\begin{Def}
Given the system of form \eqref{plant}. Problem of output regulation to zero (aka output zeroing) is the design of such state-feedback control law $u(t) = u(x)$ application of which asymptotically drives the system output to 0: $\lim_{t\to\infty} y(t) = 0$.
\end{Def}

The output zeroing problem of affine nonlinear systems can be solved using mentioned above feedback linearization method. The main idea of the method consists in the transformation using a nonlinear feedback nonlinear system $N: u(t) \mapsto y(t)$ to the linear one $L: v(t) \mapsto y(t)$ with the same outputs $y$, but new inputs $v$. After that, the resulting linear system $L$ can be controlled by means of linear control theory.

Suppose that a control problem of $N$ can be in principle solved, i.e. there is exists a satisfying input signal $u^*(t)$, which gives the output response $y^*(t)$. The essence of problems in feedback linearization comes from that the response $y^*(t)$ may not be in any way reproduced by system $L$ which is obtained after linearization.

The simplest specific example is the system $\dot x = u$, $y = h(x) = x(x^2 - 1) + 1$, $x(0) = 1$ for which the problem of output zeroing $y \to 0$ is needed to solve.

If the system under consideration was a constant relative degree, the use of control $v = -y$ after feedback linearization would give the output trajectory of $ y (t) = \exp (-t) $, which is everywhere decreasing $\dot y(t) < 0$.

In this case, the nonlinearity $y = h(x)$ has two limit points $x^{\circ}_{1,2} = \pm 1/\sqrt{3}$, in which $h'_x(x^{\circ}_{1,2}) = 0$. Any trajectory $y(t)$, that connects $y(0) = 1$ with $y(T) = 0$ passes sequentially through the points $y^{\circ}_{1} = h(3^{-1/2})$ and $y^{\circ}_{2} = h(- 3^{-1/2})$, and besides $y^{\circ}_{2} > y^{\circ}_{1}$. Hence, any trajectory $y(t)$ on the interval $(0, t_1)$ should decrease with time (Figure 1), on the interval $(t_1, t_2)$ increase, and in the interval again decrease. Such a trajectory is not reproducible using the feedback linearization.

\begin{center}
\input{Fig_0.TpX}

Figure 1. Output trajectories of linearized system with constant relative degree and system with $y = h(x) = x(x^2 - 1) + 1$
\end{center}

The behavior of the system in Figure 1 can be interpreted as follows: in the intervals $(0, t_1)$ and $(t_2, T)$ the system can be linearized in the usual manner and presented in the form $\dot y = v$. On the interval $(t_1, t_2)$ system behavior differs from the original, and the trajectory need to move in the opposite direction from the $y = 0$, which is the same as control of system $\dot y = - v $. A similar situation arises in numerical methods for finding roots and optimization of functions with singularities, where in order to achieve optimum or find a root motion in the direction opposite to predicted by Newton's method is needed. We can use the parameter $\lambda \in [0,1]$ to indicate the motion direction. Increasing of parameter $\dot \lambda> 0 $ corresponds to the movement of $y(t)$ in the direction to the desired setpoint $y = 0$, and parameter decreases $\dot \lambda < 0$ in the opposite movement. The points of direction change $ \dot \lambda (t) = 0 $ correspond to overcoming the singularities of $h(x)$. In fact, this idea is the basis of the approach proposed below.

\section{Background}

In this section we present known facts needed to understand the main result. Finally, we come to the conclusion that the numerical homotopy methods can be used not only for solving nonlinear equations, but also for control of nonlinear affine systems. Here and below we will always consider a setpoint tracking problem.

\subsection{Feedback linearization}

\begin{Def} 
MIMO nonlinear system has relative degree $r_j$ for output $y_j$ in $\mathfrak{S} \subseteq \R^n$ if at least for one function $g_i$ is true

\begin{equation}
\Lie_{g_i} \Lie_f^{r_j-1} h_j \ne 0
\end{equation}

where $\Lie_f \lambda = \frac{\partial \lambda(x)}{\partial x} f(x) = \sum_{i=1}^n \frac{\partial \lambda(x)}{\partial x_i} f_i(x)$ is a Lie derivative of function $\lambda$ along a vector field $f$. 
\end{Def} 

It means that at least one input $u_k$ influences to output $y_j$ after $r_j$ integrations.

Number $r = \sum_{i=1}^m r_j$ is called as the total relative degree of system. If $ r = n $ and matrix 

\begin{equation}\label{decmatrix}
A(x) = \begin{pmatrix} \Lie_{g_1} \Lie_f^{r_1-1} h_1(x) & \cdots & \Lie_{g_m} \Lie_f^{r_1-1} h_1(x) \\ \vdots & \cdots & \vdots \\ \Lie_{g_1} \Lie_f^{r_m-1} h_m(x) & \cdots & \Lie_{g_m} \Lie_f^{r_m-1} h_m(x) \end{pmatrix}
\end{equation}

is full rank, then the original dynamical system \eqref{plant} in $\mathfrak{S}$ equivalent to system:

\begin{equation}\label{affine_out}
y^{(r_j)}_j = \Lie_f^{r_j} h_j + \sum_{i=1}^m \Lie_{g_i} \Lie_f^{r_j-1} h_j \cdot u_i = B(x) + A(x) \cdot u
\end{equation}

The nonlinear feedback

\begin{equation}\label{nltrans}
u = A(x)^{-1} [ v - B(x) ]
\end{equation}

converts in subspace $\mathfrak{S}$ original dynamical system \eqref{plant} to linear:

\begin{equation}\label{lin}
y^{(r_j)} = v_j
\end{equation}

Control of a nonlinear system \eqref{plant} consists of two feedback loops, one of which implements a linearizing transformation \eqref{nltrans}, second one controls the system \eqref{lin} by any known method of linear control theory.

A significant drawback, which limits the applicability of the feedback linearization in practice is requirement of relative degree $r$ constancy and  full-rank of matrix $A(x)$ in the whole phase space $\mathfrak{S}$.

\subsection{Numerical continuation method}

Let it is necessary to solve system of the nonlinear equations

\begin{equation}\label{nleq}
\phi(\xi) = 0
\end{equation}

where $\phi: \R^m \to \R^m$ is vector-valued smooth nonlinear function.

Lets $\Omega \subset \R^m$ is open set and $C(\bar \Omega)$ is set of continuous maps from its closure $\bar \Omega$ to $\R^m$. Functions $F_0, F_1 \in C(\bar \Omega)$ are homotopic (homotopy equivalent) if there exists a continuous mapping

\begin{equation}\label{homotopy}
H : \bar \Omega \times [0,1] \to \R^m
\end{equation}

that $H(\xi,0) = F_0(\xi)$ and $H(\xi,1) = F_1(\xi)$ for all $\xi \in \bar \Omega$. It can be shown \cite{9} that the equation $H(\xi, \lambda) = 0$ has solution $(\xi, \lambda)$ for all $\lambda \in [0,1]$. The objective of all numerical continuation methods is tracing of implicitly defined function $H(\xi, \lambda) = 0$ for $\lambda \in [0,1]$.

Lets $H: \mathcal{D} \to \R^m$ is $C^1$-continuous function on an open set $\mathcal{D} \subset \R^{m+1}$, and the Jacobian matrix $DH(\xi,\lambda)$ is full-rank $\operatorname{rank}DH(\xi,\lambda) = m$ for all $(\xi,\lambda) \in \mathcal{D}$. Then, for all $(\xi,\lambda) \in \mathcal{D}$ exists a unique vector $\tau \in \R^{m + 1}$ such as

\begin{equation}
DH(\xi,\lambda) \cdot \tau = 0, \;
\|\tau\|_2 = 1, \;
\det \begin{pmatrix} DH(\xi,\lambda) \\ \tau^T \end{pmatrix} > 0,
\end{equation}

and mapping

\begin{equation}\label{tanfunc}
\Psi: \mathcal{D} \to \R^{m+1}, \; \Psi: (\xi,\lambda) \mapsto \tau
\end{equation}

is locally Lipschitz on $\mathcal{D}$.

Function \eqref{tanfunc} specifies the autonomous differential equation

\begin{equation}\label{homode}
\frac{d}{dt} \begin{pmatrix} \xi \\ \lambda \end{pmatrix} = \Psi(\xi,\lambda), \; \xi(0) = \xi_0, \; \lambda(0) = 0
\end{equation}

which has a unique solution $\xi(t),\lambda(t)$ according to a theorem of solution existence for the Cauchy problem.

It can be shown \cite{9} that integral curve $\gamma(t) = (\xi(t),\lambda(t))$ reaches the solution point $(\xi^{*}, 1)$ where $\phi(\xi^{*}) = 0$ in finite time $t < \infty$.

In \cite{12,13} the method of numerical homotopy continuation for the time-dependent systems of nonlinear equations $\phi(\xi,t) =0$ is given.
Another variant of numerial continuation for the system of nonstationary equations described in \cite{15}.

\section{Main result}

In \cite{12} pointed that the problem state control of the affine system \eqref{plant} with relative degree of each state $r_j = 1$ associated with the solution $\xi^*(t) : \mathbb{R}_{+} \to \mathbb{R}^n$ for nonstationary nonlinear equation $\phi(\xi, t) = 0$, in which $\frac{\partial \phi}{\partial \xi} = g(z)$, $\frac{\partial \phi}{\partial t} = f(z)$. Namely, the control $u(t) = \frac {d\xi^*(t)}{dt}$ such that $\phi(\xi^*(0), 0) = x(0)$, brings the system \eqref{plant} from the state $x(0)$ to $x(T) = 0$ asymptotically $T \to \infty$. Below we give another method than that described in \cite{12} that does not generate discontinuous control trajectories near limit points.

\subsection{Numerical continuation method for nonstationary system of nonlinear equations}

Consider the solution of the vector nonstationary equation $\phi(\xi, t) = 0$, where $\phi: \R^n \to R^n$ is a smooth function. Compose parameterized simultaneously over time $t$ and parameter $\lambda$ homotopy map:

\begin{equation}\label{homfunc}
H(\xi,\lambda,t) = (1-\lambda) \cdot (\xi - \xi_0) + \lambda \cdot \phi(\xi,t)
\end{equation}

where $\xi_0$ is initial approximation to the solution.

Lets formulate and give short proofs of some assertions for the background of parameter continuation method for nonstationary system of nonlinear equations.

\begin{Asm}\label{asm_rank}
If $\xi(t)$ and $\lambda(t)$ are solution of \eqref{homfunc}, then $\operatorname{rank}DH_{\xi,\lambda}(\xi(t),\lambda(t),t) = m$. 
\end{Asm}

From this assumption it follows that along the trajectory $(\xi(t), \lambda(t))$ there is no bifurcation points in which $\operatorname{rank} DH_{\xi,\lambda}(\xi,\lambda,t) < m$ and $D_{\lambda} H \in \operatorname{im} D_{\xi} H$. In other words, the curve $(\xi(t), \lambda(t))$ is free from branches and self-intersections.

Equation $H(\xi,\lambda,t) = 0$ for $\lambda \in [0, 1]$ defines implicitly defined function $\xi(t)$, parameterized by $\lambda(t)$ and satisfies the equation obtained by differentiating \eqref{homfunc} by time

\begin{equation}\label{homfuncdt}
\dot H(\xi(t),\lambda(t),t) = \frac{\partial H}{\partial \xi} \dot \xi(t) + \frac{\partial H}{\partial \lambda} \dot \lambda(t) +  \frac{\partial H}{\partial t} = 0
\end{equation}

Denoting

\begin{equation}
\begin{gathered}
A = \left( \frac{\partial H}{\partial \xi} ~ \frac{\partial H}{\partial \lambda} \right ), \;
B = - \frac{\partial H}{\partial t}, \;
\tau = \begin{pmatrix} \dot \xi \\ \dot \lambda \end{pmatrix}
\end{gathered}
\end{equation}

\eqref{homfuncdt} can be represented as a linear matrix equation with respect to $\dot \xi$ and $\dot \lambda$:

\begin{equation}\label{hommateq}
A \tau = B
\end{equation}

\begin{Lmm} 
Equation \eqref{hommateq} always has a solution.
\end{Lmm}

\begin{Prf}
Because of $A = DH_{\xi,\lambda}(\xi,\lambda,t)$ the undetermined equation \eqref{hommateq} has a solution if and only if $\operatorname{rank}DH_{\xi,\lambda}(\xi,\lambda,t) = m$, what is based on the assumption 1. $\square$
\end{Prf}

\begin{Lmm} \label{lmm_solcomps}
All the solutions of \eqref{hommateq} can be represented as $\tilde \tau = \alpha \cdot \tau + \bar \tau$, where $\bar \tau = A^{+} B$, $\tau \in \operatorname{ker} A$, $\alpha \in \R$. 
\end{Lmm}

\begin{Prf}
It is quite obvious, since $\operatorname{dim} \operatorname{ker} A = 1$ and all the null-space of $A$ can be parametrized by scalar variable $\alpha \in \R$, and the solution space $W$ inhomogeneous equation of the form \eqref{hommateq} is defined as $W = \{ A^{+} B \} \oplus \operatorname{ker} A$. $\square$
\end{Prf}

Let's prove assertion of Lipschitz maps $\Psi: \mathcal{D} \to \R^{m+1}$, which will need further.

\begin{Thm} \label{thm_ode}

Let $H: \mathcal{D} \to \R^m$ is $C^1$-continuous function on an open set $\mathcal{D} \subset \R^{m+2}$, and the Jacobian matrix $A$ is full-rank $\operatorname{rank} A = m$ for all $(\xi,\lambda,t) \in \mathcal{D}$. Then for each $(\xi,\lambda,t) \in \mathcal{D}$ exists a unique vector $\tilde \tau \in \R^{m + 1}$, such that

\begin{equation}
\begin{gathered}
\tilde \tau = \alpha \cdot \tau + \bar \tau, \; \bar \tau = A^+ B, \\
A \cdot \tau = 0, \;
\|\tau\|_2 = 1, \;
\det \begin{pmatrix} A \\ \tau^T \end{pmatrix} > 0, \; \alpha = \operatorname{const} \in \R_{+},
\end{gathered}
\end{equation}

and map

\begin{equation}\label{tanfunc}
\tilde \Psi: \mathcal{D} \to \R^{m+1}, \; \tilde \Psi: (\xi,\lambda,t) \mapsto \tilde \tau
\end{equation}

is locally Lipschitz on $\mathcal{D}$. 

\end{Thm}

\begin{Prf}

Vector $\tilde \tau = \alpha \cdot \tau + \bar \tau$ defined as the sum of two components, one of which $\tau$ is known (initial value problem 2.1.9 in \cite{9}), that it is a Lipschitz function on $\mathcal{D}$. Hence, it is necessary to prove that $\bar \tau = A^+ B = (DH_{\xi, \lambda})^+DH_{t}$is Lipschitz. The uniqueness of the $\bar \tau$ follows from the uniqueness of the Moore-Penrose matrix pseudo-inversion. 

We can assume that $DH$ is Lipschitz on $\mathcal{D}$ with a constant $\gamma$, which leads to existence and boundedness of the second derivatives $H$. Since $\operatorname{rank} A = m$, the pseudo-inversion function $A^+$ continuous and differentiable \cite{11}. Hence the product $A^+ B$ is Lipschitz, because its components are Lipschitz. $\square$

\end{Prf}

Function $\tilde \Psi$ specifies the autonomous differential equation

\begin{equation}\label{homtode}
\frac{d}{dt} \begin{pmatrix} \xi \\ \lambda \end{pmatrix} = \tilde \Psi(\xi,\lambda,t), \; \xi(0) = \xi_0, \; \lambda(0) = 0
\end{equation}

which has a unique solution $\xi(t),\lambda(t)$ according to a theorem of existence and uniqueness of solutions of the Cauchy problem.

\begin{Thm}

The set $H^{-1}(\{ 0 \})$ is simply connected.

\end{Thm}

\begin{Prf}

By theorem 2.1 from \cite{20}, if for the map $F: \R^n \to \R^k$, $k \le n$ true that

\begin{equation}
\operatorname{sup}\{ \| (DF(x) D^T F(x))^{-1} \| ~|~ x \in \R^n \} < \infty
\end{equation}

then $F^{-1}(0)$ connected submanifold of dimension $(n-k)$ в $\R^n$.

In accordance with the assumption \ref{asm_rank}, $\rank DH = m$, therefore $\rank D^T H = m$, $\rank DH \cdot D^T H = m$, inversion of the matrix $(DH \cdot D^T H)^{-1}$ defined and its norm is bounded. $\square$

\end{Prf}

\begin{Thm}

A necessary condition for the existence of integral curve $(\xi(t), \lambda(t))$ for \eqref{homtode} that connects the points $(\xi_0, 0)$ and $(\xi^*, 1)$ is $\alpha > 0$.

\end{Thm}

\begin{Prf}

Based on lemma \ref{lmm_solcomps} and theorem \ref{thm_ode} it is obvious that the Cauchy problem of the form \eqref{homtode} has a unique solution $(\xi(t), \lambda(t))$, that satisfies the equation \eqref{homfunc} for each fixed $t = \const$: $H(\xi(t),\lambda(t),t) = 0$.

To determine the sign of $\alpha$ consider the behavior of the curve $\gamma(t) = (\xi(t), \lambda(t))$ near $t = 0$. Since 

\begin{equation}\label{AB_explicit}
\begin{gathered}
A = \left( \lambda \frac{\partial \phi}{\partial \xi} + (1-\lambda) E ~~ \phi - \xi \right ), \;
B = - \lambda \frac{\partial \phi}{\partial t}
\end{gathered}
\end{equation}

then near $t = 0$ equation $\eqref{homfunc}$ behaves as a stationary with $B = 0$. Let's show that $\alpha \ne 0$. If $\alpha = 0$, then at time $t = 0$, from lemma \ref{lmm_solcomps} and \eqref{AB_explicit} it follows that $\dot \xi(0) = 0$, $\dot \lambda(0) = 0$, thereof $\dot \lambda(t) = 0$ for each $t$ and condition $\lambda = 1$ never attainable. Since the near $t = 0$ equation $\eqref{homfunc}$ stationary and the solution is determined by the conventional method of parameter continuation \cite{9}, then orientation of vector $\tau$ must be agreed with the condition $\begin{pmatrix} A \\ \tau^T \end{pmatrix} > 0$. Hence $\alpha > 0$. $\square$

\end{Prf}

Now we state and give a short proof of the assumption about the behavior of the solution curve $\gamma(t) = (\xi(t),\lambda(t))$ of \eqref{homtode}. 

\begin{Thm}

There exists a number $\alpha_0 \in \R$, that the integral curve $\gamma(t) = (\xi(t),\lambda(t))$ for equation \eqref{homtode} with $\alpha > \alpha_0$ has finite length between the points $(\xi_0, 0)$ and $(\xi^*, 1)$.

\end{Thm}

\begin{Prf}

To prove this we consider the structure of the right side \eqref{homtode}. Since the constant $\alpha$ can be selected arbitrarily large, then the term $\bar \tau$ can be neglected and may be written: 

\begin{equation}
\tilde \tau = \alpha \cdot \tau + \bar \tau \xrightarrow[\alpha \to \infty]{} \alpha \cdot \tau
\end{equation}

By theorem \ref{thm_ode} the function $\bar \tau$ is Lipschitz on $\mathcal{D}$, then it follows automatically that $\bar \tau$ is bounded. Then one can always choose a finite $\alpha$ so that $\tilde \tau = \alpha \cdot \tau + \epsilon$, $\epsilon \ll \alpha \cdot \tau$.

Since $\alpha$ is a finite number, then $\alpha \cdot \tau$ as right side of \eqref{homtode} satisfies the well-known results on the finiteness of the solution trajectory (lemma 2.1.13 and theorem 2.1.14 in \cite{6}). Hence, for an appropriate choice $\alpha > \alpha_0$ curve $(\xi(t),\lambda(t))$ has no limit points, and is diffeomorphic to the line, i.e. has finite length between the $\lambda_0 = 0$ и $\lambda_1 = 1$. $\square$

\end{Prf}

The last theorem indicates that the parameter $\alpha$ is another one degree of freedom in designing the controller. The larger this constant, the faster the solution arrives to the $\lambda = 1$, but numerical integration becomes more stiff.

\subsection{Homotopy continuation for nonlinear affine systems}

Let's associate with the plant \eqref{plant} linear dynamics system with $m$ inputs $u$, $n$ states $z$, $m$ outputs $\eta$ and with the same relative degries $r_i$ for outputs such as in \eqref{plant}

\begin{equation}\label{linsys}
\dot z = A z + B u, \; \eta = C z, \; \frac{d^{(r_i)}}{d t^{(r_i)}}\eta = u_i.
\end{equation}

Following equation is the homotopy mapping that links the outputs dynamics of the system \eqref{plant} and \eqref{linsys}:

\begin{equation}
H = (1 - \lambda) \cdot \eta + \lambda \cdot y = 0.
\end{equation}

By definition of the relative degree of output, each component $H_i$ should be differentiated $r_i$ times with respect to $t$ until it becomes an explicit function of any input $u$. We obtain after differentiation:

\begin{equation}\label{difform_H_i_pre}
\begin{gathered}
H_i^{(r_i)} = - \sum_{k=1}^{r_i-1} C_{r_i}^k \eta_i^{(r_i-k)} \lambda^{(k)} + (1 - \lambda) u_i + (y_i - \eta_i) \lambda^{(r_i)} + \sum_{k=1}^{r_i-1} C_{r_i}^k y_i^{(r_i-k)} \lambda^{(k)} + \\
\quad + \lambda \left ( \Lie_f^{r_i} h_i + \sum_{k=1}^m \Lie_{g_k} \Lie_f^{r_i-1} h_i \cdot u_k \right ) = 0,
\end{gathered}
\end{equation}

that gives:

\begin{equation}\label{difform_H_i}
H^{(r_i)} = \mathcal{A}_{i,1}(x,z,\Lambda) \cdot u + \mathcal{A}_{i,2}(x,z,\Lambda) \cdot \lambda^{(r_i)} + \mathcal{B}_i(x,z,\Lambda),
\end{equation}

where $\Lambda = (\lambda, \dot \lambda, \ddot \lambda, ... , \lambda^{(r_i-1)})$, $C_n^k$ are binomial coefficients.

Considering all of the components $H_i$ after differentiation according to the relative degrees of outputs $r_i$ it is possible to write an algebraic condition that specifies a continuous deformation of system \eqref{linsys} to \eqref{plant}.

\begin{equation}\label{H_all}
\bar H = \mathcal{A}_{1}(x,z,\bar \Lambda) \cdot u + \mathcal{A}_{2}(x,z,\bar \Lambda) \cdot \lambda^{(r_{max})} + \mathcal{B}(x,z,\bar \Lambda) = 0,
\end{equation}

where $r_{max} = \max\{ r_i \}$, $\bar \Lambda = (\lambda, \dot \lambda, \ddot \lambda, ... , \lambda^{(r_{max}-1)})$, 

If $\mathcal{B} = 0$, that is corresponds to the case $r_i = 1$, $f(x) = 0$, any known method of parameter continuation (like predictor-corrector method \cite{9}) can determine the trajectory $u(t), \lambda(t)$ such as $ \begin{pmatrix} u(t) \\ \dot \lambda(t) \end{pmatrix} = \tau(t) $, where $\tau(t)$ is tangent vector to the implicit curve $H=0$, which is obtained from the linear matrix equation $\mathcal{A} \tau = 0$. This equation has infinitely many solutions as there are $n$ conditions and $n+1$ variables. In order to uniquely identify $u(t)$ and $\dot \lambda(t)$ an additional condition $ \|\tau \| = 1$ for length normalization of the vector $\tau$ needed. In addition, to select the correct direction of the $\tau$, imposed a condition of its positive orientation relative to the surface $H$, given in the form of inequality $ \det \begin{pmatrix} \mathcal{A} \\ \tau^T \end{pmatrix} > 0 $.

Considering the general case $\mathcal{B} \ne 0$, tangent vector $\tau(t)$ needs to be augmented by term for notstationarity compensating. Connected path $(u(t), \lambda(t))$ for $t \in [0, T)$ starting at point $(u_0, 0)$ such as 

\begin{equation}
\lim_{t \to T} \lambda(t) = 1 , \; \bar H(u(t),\lambda(t),t) = 0,
\end{equation}

can be generated from \eqref{H_all} as follows:

\begin{equation}\label{Sol}
\begin{gathered}
\begin{pmatrix} u \\ \lambda^{(r_{max})} \end{pmatrix} = \alpha \cdot \tau + \bar \tau, \\
\bar \tau = \mathcal{A}^+ \mathcal{B}, \\
\mathcal{A} \cdot \tau = 0, \; \|\tau\|_2 = 1, \; \det \begin{pmatrix} \mathcal{A} \\ \tau^T \end{pmatrix} > 0,
\end{gathered}
\end{equation}

with the additional condition

\begin{equation}\label{ExistCond}
\operatorname{rank}\mathcal{A}(x,z,\bar \Lambda) = m,
\end{equation}

where $\alpha > 0, ~ \alpha \in \R$ is a scalar constant, $A^+$ is Moore-Penrose inverse of matrix $A$.

Condition \eqref{ExistCond} is a standard assumption when using parameter continuation method, which corresponds to the possible existence of limit points of trajectories $(u(t), \lambda(t))$ at which $\mathcal{A}_1 \notin \operatorname{im} \mathcal{A}_2$, and the absence of bifurcation points. At the same time in some regions of phase space $X \times Z$ may be a situation where $\operatorname{rank} \mathcal{A}_1(x,z,\bar \Lambda) < m$, in that case, the system can not be linearized by the feedback, but the proposed method is applicable. Overcoming the bifurcation points, in which is observed $\mathcal{A}_1 \in \operatorname{im} \mathcal{A}_2$, also possible within the known approaches for the numerical parameter continuation (e.g., using the Lyapunov-Schmidt decomposition \cite{9}).

Equation \eqref{H_all} specifies the state feedback, and \eqref{Sol} specifies dynamics of the controller. According to \eqref{difform_H_i_pre}, equations for $\mathcal{B}$ and $\mathcal{A}_2$ depend explicitly on the $y$, and thus implements the output feedback.

\subsection{Switching strategies for nonregular feedback linearization}

All plants in practice are subject to a variation of parameters. The variant of control offered in the previous section is definitely sensitive to parametric uncertainties in a control object. On the other hand, in the feedback linearization control the parameter variation in the plant can be compensated by the controller for the linearized system \cite{16}. Let's consider a hybrid method that combines the possibility of applying an external control loop and resistant to change in the relative degree of the system.

System in form \eqref{difform_H_i_pre} with output $H$ can be linearized by feedback if consider evalution of parameter $\lambda$ as observable internal dynamics. In case if we fix $\lambda^{(r_{max})} = \const$, $\lambda^{(r_{max})} \in \{-1, 1\}$, then we obtain from \eqref{difform_H_i_pre} following affine nonlinear system:

\begin{equation}\label{H_fblin_}
\begin{gathered}
\bar H = F(x,z,\bar \Lambda) + G(x,z,\bar \Lambda) \cdot u \\
F(x,z,\bar \Lambda) = \mathcal{A}_{2}(x,z,\bar \Lambda) \cdot \lambda^{(r_{max})} + \mathcal{B}(x,z,\bar \Lambda) \\
G(x,z,\bar \Lambda) = \mathcal{A}_{1}(x,z,\bar \Lambda)
\end{gathered}
\end{equation}

It's possible to write a nonlinear coordinate transformation $v \mapsto u$ transforming a nonlinear system \eqref{H_fblin_} to linear one $H^{(r)} = v$

\begin{equation}\label{H_fblin_trans_}
u = [\mathcal{A}_{1}(x,z,\bar \Lambda)]^{-1} \left ( v - \mathcal{A}_{2}(x,z,\bar \Lambda) \cdot \lambda^{(r_{max})} - \mathcal{B}(x,z,\bar \Lambda) \right ),
\end{equation}

Switching strategy can be described as follows: we start whith conventional feedback linearization of \eqref{H_fblin_}, then in areas $\bar{ \mathfrak{S}}$ where $\det G(x,z,\bar \Lambda) \approx 0$ and feedback linearization is not possible, it is necessary to switch control to the parameter continuation method implemented with \eqref{Sol}.

It should be noted that when $v = 0$ control \eqref{H_fblin_trans_} is a special case of parameter continuation strategy that applied far from limit points. If $\lambda^{(r_{max})} \ne 0$, then control \eqref{H_fblin_trans_} is generated by equations \eqref{Sol} with following time-dependent scaling:

\begin{equation}
\alpha = \frac{\sign(\hat \lambda^{(r_{max})} ) - \bar \lambda^{(r_{max})}}{\hat \lambda^{(r_{max})}}, \; \begin{pmatrix} \hat u \\ \hat \lambda^{(r_{max})} \end{pmatrix} = \tau, \; \begin{pmatrix} \bar u \\ \bar \lambda^{(r_{max})} \end{pmatrix} = \bar \tau
\end{equation}

Since the degeneracy of matrix $\mathcal{A}_{1}$ in feedback linearization control always leads to increasing to infinity at least one input in $u$, practical way to perform switching between control strategies is consider event of inputs saturation when $\max{|u_i|} > u_{max}$. Input signals are always limited $|u| \le u_{max}$ in the real world applications. This yields the following algorithm of the hybrid feedback linearization: we start with positive sign $s = 1$ of $\lambda^{(r_{max})} = s$ and conventional feedback linearization, then in case of saturation of inputs we switch to homotopy continuation procedure, and after return of all input signals to its limits we flipped sign $s := -s$ and switch back to feedback linearization. Formal procedure for calculating the control actions can be represented as follows

\begin{algorithmic}

\State $u := [\mathcal{A}_{1}(x,z,\bar \Lambda)]^{-1} \left ( v - \mathcal{A}_{2}(x,z,\bar \Lambda) \cdot s - \mathcal{B}(x,z,\bar \Lambda) \right )$

\If {$\max{|u_i|} > u_{max}$}
    \State $\begin{pmatrix} \hat u \\ \hat \lambda^{(r_{max})} \end{pmatrix} = \tau, \; \mathcal{A} \cdot \tau = 0, \; \|\tau\|_2 = 1, \; \det \begin{pmatrix} \mathcal{A} \\ \tau^T \end{pmatrix} > 0$
	\State $\begin{pmatrix} \bar u \\ \bar \lambda^{(r_{max})} \end{pmatrix} = \mathcal{A}^+ \mathcal{B}$	
	\State $k_{\lambda} := (\sign(\hat \lambda^{(r_{max})} ) - \bar \lambda^{(r_{max})}) / \hat \lambda^{(r_{max})}$
	\State $k_u := (\sign(\max{|u_i|}) u_{max} - \max{|u_i|}) / \max{|u_i|}$
	\If {$k_u > k_{\lambda}$}
		\State $\hat u := \hat u \cdot k_{\lambda}$
		\State $s := \sign (\hat \lambda^{(r_{max})})$
		\State $\hat \lambda^{(r_{max})} := s$
	\Else {$k_u > 1$}
		\State $\hat u  := \hat u  \cdot k_u$
		\State $\hat \lambda^{(r_{max})} := \hat \lambda^{(r_{max})} \cdot k_u$
	\EndIf
	\State $u = \hat u + \bar u$, $\lambda^{(r_{max})} = \hat \lambda^{(r_{max})} + \bar \lambda^{(r_{max})}$
\EndIf

\end{algorithmic}

Since the set of regions $\bar{ \mathfrak{S}}$ forms a compact space, using of this algorithm prodice almost linearized system. The situation where $x \in \bar{ \mathfrak{S}}$ can be considered as a perturbation action in the output $H$.

\section{Applications}

\subsection{One illustrative example}

Consider following abstract example of MIMO system, that changes its relative degree in the state space

\begin{equation}\label{plant_ex_1}
\begin{gathered}
\dot x_1 = u_1 + x_2^3 \\
\dot x_2 = u_2 + x_1^3 \\
y_1 = x_1^3 - x_1 + 1 \\
y_2 = x_2^4 \cos(2 x_2)
\end{gathered}
\end{equation}

with initial conditions $x(0) = (1,1)^T$. We need to solve the problem of output zeroing $y \to 0$. Let's suppose that input signals constrained by inequality $| u_i | \le 20$.

Differentiating the outputs, we obtain

\begin{equation}\label{plant_ex_1_dy}
\begin{gathered}
\dot y_1 = \left (3 x_1^2 - 1 \right ) \cdot (u_1 + x_2^3) = g_{11} u_1 + f_1 \\
\dot y_2 = \left ( 4 x_2^3 \cos(2 x_2) - 2 x_2^4 \sin(2 x_2) \right ) \cdot (u_2 + x_1^3) = g_{22} u_2 + f_2
\end{gathered}
\end{equation}

Obviously, the system in interval $x \in [0,1]^2$ can not be completely linearized by the feedback, because there are exists such $x^*$ that $g_{11}(x^*) = 0$ or $g_{22}(x^*) = 0$.

Let's associate with \eqref{plant_ex_1} linear system of a form

\begin{equation}\label{plant_ex_1_lin}
\begin{gathered}
\dot \eta_1 = u_1, \; \dot \eta_2 = u_2
\end{gathered}
\end{equation}

with initial conditions $\eta(0) = (0,0)^T$

According to the equation \eqref{difform_H_i_pre} we obtain for $\dot H = 0$ the following

\begin{equation}\label{H_all_ex_1}
\mathcal{A}_{1} = \begin{pmatrix} g_{11} & 0 \\ 0 & g_{22} \end{pmatrix} \cdot u + (1-\lambda) E, \; \mathcal{A}_{2} = y - \eta, \; \mathcal{B} = -\begin{pmatrix} f_1 \\ f_2 \end{pmatrix}
\end{equation}

The model in Simulink to control the system \label{plant_ex_2} shown in figure 2. Modeling results are shown on figures 3-4.

\begin{center}
\includegraphics[width=1.0\textwidth]{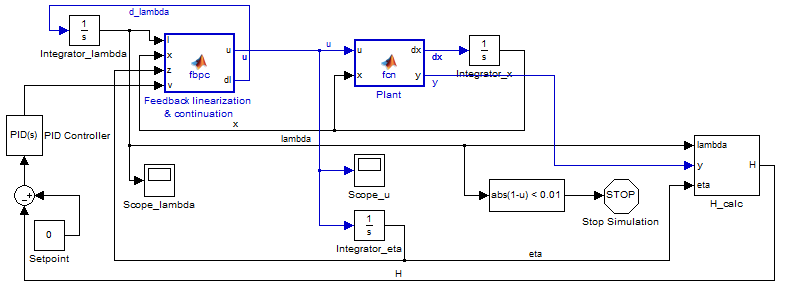}

Figure 2. Simulink model.
\end{center}

\begin{center}
\includegraphics[width=0.85\textwidth]{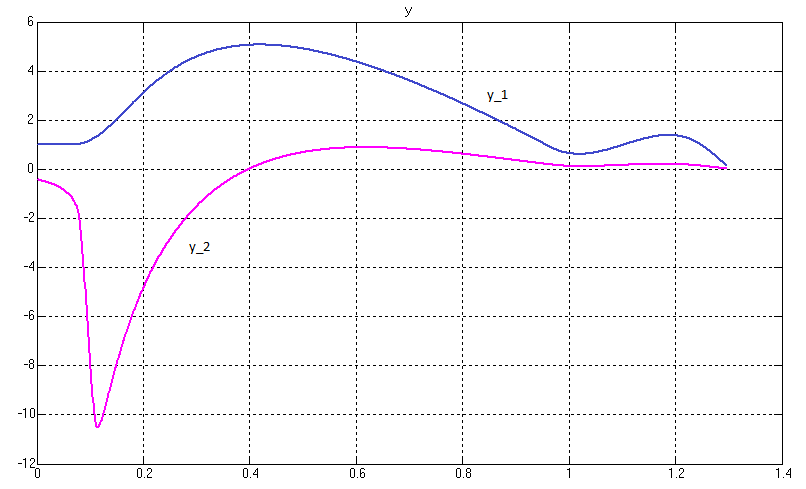}

Figure 3. Output response.
\end{center}

\begin{center}
\includegraphics[width=0.85\textwidth]{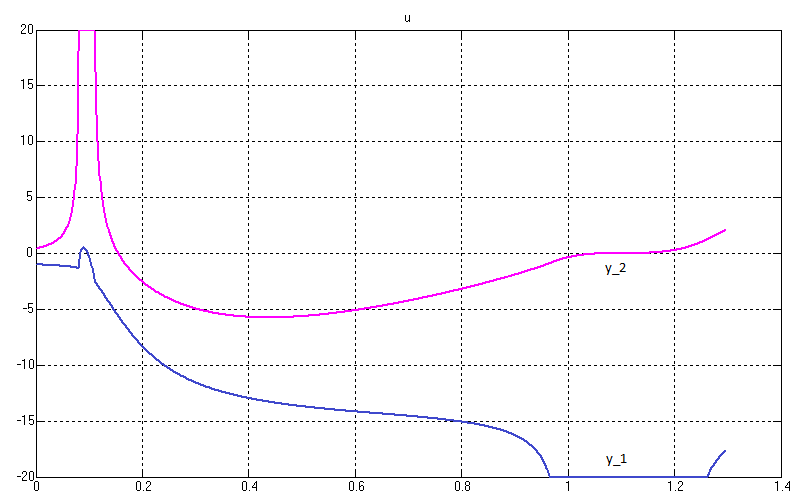}

Figure 4. Input controls.
\end{center}

\subsection{Three phase induction motor control}

Three-phase asynchronous motor is a famous example of a system that can not be linearized by state feedback \cite{17}. Consider the application of the proposed method to control the speed and flux linkage of the motor. For modeling of the electric motor in the state space, we strictly follow the material of the paper \cite{18}.

Let's consider the reduced fourth-order state-space model of induction modor:

\begin{equation}\label{reduced_plant}
\begin{gathered}
\dot \omega = p^2 \frac{ M_{sr} \phi_r i_{sq}}{J L_r} - \frac{p T_m}{J} \\
\dot \phi_r = - \tau_r^{-1} \phi_{rd} + \tau_r^{-1} M_{sr} i_{sd} \\
\dot i_{sd} = \beta \tau_r^{-1} \phi_r - \tau_1^{-1} i_{sd} + \omega_s i_{sq}    +    \frac{V_{sd}}{L_1} \\ 
\dot i_{sq} = -\beta \omega \phi_r - \tau_1^{-1} i_{sq} - \omega_s i_{sd}    +    \frac{V_{sq}}{L_1}
\end{gathered}
\end{equation}

with this kind of parametrization:

\begin{equation}\label{plants_params}
\begin{gathered}
\tau_r = \frac{L_r}{R_r} , \; \mu = p^2 \frac{M_{sr}}{J L_r} , \; \beta = \frac{M_{sr}}{L_r L_1} \\
L_1 = L_s - \frac{M_{sr}^2}{L_r}, \; R_1 = R_s + R_r \left ( \frac{M_{sr}}{L_r} \right )^2, \; \tau_1 = \frac{L_1}{R_1}
\end{gathered}
\end{equation}

where $i_{sd}$ and $i_{sq}$ are respectively the stator currents projections on the $(d, q)$ axis reference frame, $\phi_r$ is a rotor fluxe, $L_s$ and $L_r$ are the stator and rotor self-inductances and $M_{sr}$ is the mutual inductance. 

The electromagnetic torque developed by the motor is expressed in terms of rotor fluxes and stator currents as:
\begin{equation}\label{plant_out}
T_{em} = p \frac{M_{sr}}{L_r}(i_{sq} \phi_r )
\end{equation}

where $p$ is a number of pole pairs.

Synchronous rotor angular speed $\omega_s$ can be expressed as

\begin{equation}\label{sync_speed}
\omega_s = \omega + \frac{M_{sr} i_{sq}}{\tau_r \phi_r}
\end{equation}

The outputs to be controlled are the mechanical speed $y_1 = \omega / p$ and the square of the rotor flux magnitude $y_2 = \phi_r^2$. State variables are stator currents $(i_{sd}, i_{sq})$, the rotor fluxes $(\phi_{rd}, \phi_{rq})$ and the rotor angular speed $\omega$. Control variables are stator voltages $V_{sd}$ and $V_{sq}$.

First differentiation of outputs yields

\begin{equation}\label{d_reduced_plant}
\begin{gathered}
\dot y_1 = \dot \omega / p  = \mu \phi_r i_{sq} / p - T_m/J \\
\dot y_2 = -2 \tau_r^{-1} \phi_r^2 + 2 M_{sr} \tau_r^{-1} \phi_r i_{sd}
\end{gathered}
\end{equation}

After second differentiation of outputs finally inputs appeared:

\begin{equation}\label{dd_reduced_plant}
\begin{gathered}
\begin{pmatrix} \ddot y_1 \\ \ddot y_2 \end{pmatrix} = \begin{pmatrix} b_1 \\ b_2 \end{pmatrix} + \begin{pmatrix} 0 & a_{12} \\ a_{21} & 0 \end{pmatrix} \\
b_1 = -\mu \phi_r (i_{sq} \tau_1^{-1} + i_{sq} \tau_r^{-1} + \omega_s i_{sd}) + \mu \tau_r^{-1} M_{sr} i_{sd} i_{sq} - \mu \beta \omega \phi_r^2 \\
b_2 = \frac{2} {\tau_r^{2}} (2 + \beta M_{sr}) \phi_r^2 - \left (3 \frac {M_{sr}}{\tau_r} + \frac {M_{sr}}{\tau_1} \right) \frac{2 i_{sd} \phi_r}{\tau_r} + 2 M_{sr} \omega_s \frac{i_{sq} \phi_r}{\tau_r} + 2 M_{sr}^2 \frac{i_{sd}^2}{\tau_r^2} \\
a_{12} = {\mu}{L_1 \phi_r} \\
a_{21} = \frac{2 M_{sr} \phi_r}{\tau_r L_1}
\end{gathered}
\end{equation}

Plant model in the form of \eqref{dd_reduced_plant} can be linearized by feedback when $\phi_r \ne 0$, after nonlinear transformation we will have

\begin{equation}\label{lin_plant}
\ddot y = v
\end{equation}

In \cite{18} to control the \eqref{lin_plant} proportional-differential (PD) controller used, which in practice has a number of fundamental problems of reducing the stability to noise in the feedback.

In this paper we propose a different approach to the control of \eqref{dd_reduced_plant}, based on two feedback loops: the internal to stabilize the current $i_{sd}, i_{sq}$ and the external to control outputs $y_1, y_2$. As a result, only proportional-integral (PI) controllers are used and structure of the system resembles a classical FOC-control with the only difference being that the output of each PI controller is passed through an appropriate nonlinear transformation of coordinates.

Parameter continuation is used only in the outer control loop, the inner loop is implemented with a current $i_{sd}, i_{sq}$ decoupling by coordinate transformation \cite{21}.

Inner loop for current stabilization is implemented using a nonlinear feedback through which control signals $\nu_{sd}$, $\nu_{sq}$ are passed

\begin{equation}\label{nltrans_current}
\begin{gathered}
u_{sd} = L_1 (\nu_{sd} -  \omega_s i_{sq}) \\ 
u_{sq} = L_1 (\nu_{sq} + \beta \omega \phi_r + \omega_s i_{sd})
\end{gathered}
\end{equation}

which gives the decoupled linear dynamics of the currents

\begin{equation}\label{current_plant}
\begin{gathered}
\dot i_{sd} = \nu_{sd} + \beta \tau_r^{-1} \phi_r - \tau_1^{-1} i_{sd} \\ 
\dot i_{sq} = \nu_{sq} - \tau_1^{-1} i_{sq} 
\end{gathered}
\end{equation}

Control of \eqref{current_plant} can be achieved with a simple PI controller

\begin{equation}\label{current_p}
\nu = K_p (i^{ref} - i) + K_i \int_0^t (i^{ref}(\tau) - i(\tau) ) d \tau
\end{equation}

where $i^{ref}$ is a current setpoint for the corresponding axis.

With corresponding adjustment of coefficients $K_p$ and $K_i$ can be achieved fast regulation and exact match $i_{sd} \approx i_{sd}^{ref}$, $ i_{sq} \approx i_{sq}^{ref}$, which allows ignore the dynamics of the current regulation \cite{21}.

Let us turn to the outer loop to control the outputs of $y_1 = \phi_r^2$ and the mechanical speed $y_2 = \omega / p$, whose dynamics is given by the equation \eqref{d_reduced_plant}. Let's associate with \eqref{d_reduced_plant} linear system of a form

\begin{equation}\label{reduced_plant_lin}
\dot \eta_1 = u_1 = i_{sd}, \; \dot \eta_2 = u_2 = i_{sq}
\end{equation}

with initial conditions $\eta(0) = (0,0)^T$.

Using \eqref{difform_H_i_pre} and \eqref{difform_H_i} we can write the following equation for mixed dynamics of \eqref{d_reduced_plant} and \eqref{reduced_plant_lin}

\begin{equation}\label{plant_ex_full}
\begin{gathered}
\dot H = \lambda \dot y + u (1 - \lambda) + \dot \lambda (y - \eta) = 0 \\
\dot H_1 = \lambda (-2 \tau_r^{-1} \phi_r^2 + 2 M_{sr} \tau_r^{-1} \phi_r i_{sd}) + i_{sd} (1 - \lambda) + \dot \lambda (y_1 - \eta_1) = 0 \\
\dot H_2 = \lambda \mu \phi_r i_{sq} / p  + i_{sq} (1 - \lambda) + \dot \lambda (y_2 - \eta_2) = 0
\end{gathered}
\end{equation}

The purpose control is the asymptotical output zeroing $H_1(t) \to 0$, $H_2(t) \to 0$. Should be noted that we droped therm $T_m/J$ in equation \eqref{d_reduced_plant} for $y_2$. Для этого осуществим линеаризацию системы \eqref{plant_ex_full}.

Equation \eqref{plant_ex_full} establishes algebraic condition for the continuous deformation of the system \eqref{reduced_plant_lin} to \eqref{d_reduced_plant}. It is used to control the system in regions where feedback linearization is not possible, i.e. if $\phi_r \ne 0$. The control inputs calculated by \eqref{Sol}, which has particular form of 

\begin{equation}\label{dH_linearizaton}
\begin{gathered}
i_{sd} = \frac{2 \lambda \phi_r^2 + v_1 \tau_r + \eta \dot \lambda \tau_r - \dot \lambda \tau_r y}{\tau_r - \lambda \tau_r + 2 M_{sr} \lambda \phi_r} \\
i_{sq} = \frac{ v_2 + \eta \dot \lambda - \dot \lambda y}{\mu \lambda \phi_r / p + 1 - \lambda}
\end{gathered}
\end{equation}

with initial dynamic of paremeter $\lambda$: $\lambda(0) = 0$, $\dot \lambda(0) = 1$. Inputs $v_1$ and $v_2$ are controlled with convential PI regulators 

\begin{equation}\label{y_plant}
\begin{gathered}
v_1 = K_{p1} (y_1^{ref} - y_1) + K_{i1} \cdot e_1, \; \dot e_1 = y_1^{ref} - y_1 \\
v_2 = K_{p2} (y_2^{ref} - y_2) + K_{i2} \cdot e_2, \; \dot e_2 = y_2^{ref} - y_2
\end{gathered}
\end{equation}

When the system is far from area of singularity of linearizing transformation (i.e. when $\phi_r > 0$) control \eqref{dH_linearizaton} reducec into \eqref{nltrans_y}

\begin{equation}\label{nltrans_y}
\begin{gathered}
i_{sd} = \frac{\tau_r}{2 M_{sr} \phi_r} ( v_1 + 2 \tau_r^{-1} \phi_r^2 ) \\
i_{sq} = \frac{p}{\mu \phi_r} v_2
\end{gathered}
\end{equation}

Switching between \eqref{nltrans_y} and \eqref{dH_linearizaton} occurs based on the analysis of exceeding the boundaries $u_{max}$ of control actions according to the algorithm described in the end of section 4.

Informally, the essence of the proposed control is reduced to that the control object \eqref{d_reduced_plant} in area close to $\phi_r = 0$ is parametrically replaced to \eqref{reduced_plant_lin}. For this parametrized plant conventional feedback linearization is applied. 

\subsection{Three phase induction motor control: simulation}

This numerical experiment is conducted to simulate the start of induction motor and stabilization of speed and flux. In area $\phi_r \approx 0$ induction machine cannot be linearized by feedback and open loop controller usually used for start \cite{18}. With our parametrization its possibe to perform feedback lineraization of motor on start as well. 

The model in Simulink to control the system \label{reduced_plant} shown in figures 5-6. 

\begin{center}
\includegraphics[width=1.0\textwidth]{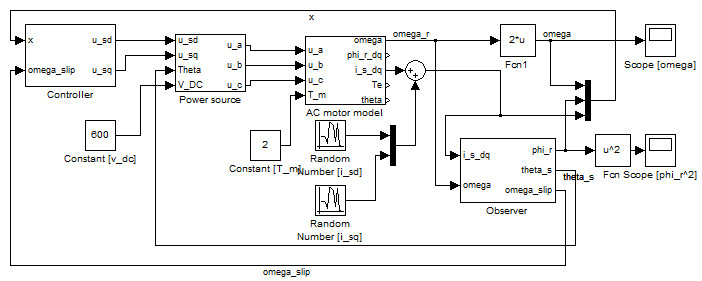}

Figure 5. Simulink model of overall system.
\end{center}

The model consists of the following hierarchical blocks:

\begin{itemize}
\item AC motor model -- model of induction motor in the form \eqref{reduced_plant},
\item Power source -- inverse Park transformation and the SVPWM signal generator,
\item Controller -- subsystem with the PI controllers and the linearizing transformations for speed control $\omega$ flux $\phi_r$,
\item Observer -- subsystem block implements the calculation of the slip speed $\omega_{slip}$ and flux $\phi_r$ from measured currents $i_{sd}$, $i_{sq}$
\end{itemize}

The controller subsystem is shown in Figure 5. Model of the controller consists of the following blocks

\begin{itemize}
\item Plant [w, phi] -- nonlinear coordinate transformation in form of \eqref{dH_linearizaton} to control the speed $\omega$ and $\phi_r$,
\item Plant [i\_s\_dq] -- nonlinear coordinate transformation in form of \eqref{nltrans_current} to control the stator currents $i_{sd}$ and $i_{sq}$,
\item PID Controller [i\_sd] -- PI-controller for stabilization of $i_{sd}$,
\item PID Controller [i\_sq] -- PI-controller for stabilization of $i_{sq}$,
\item PID Controller [phi\_r] -- PI-controller for stabilization of $\phi_r$,
\item PID Controller [omega] -- PI-controller for stabilization of $\omega$,
\end{itemize}

\begin{center}
\includegraphics[width=1.0\textwidth]{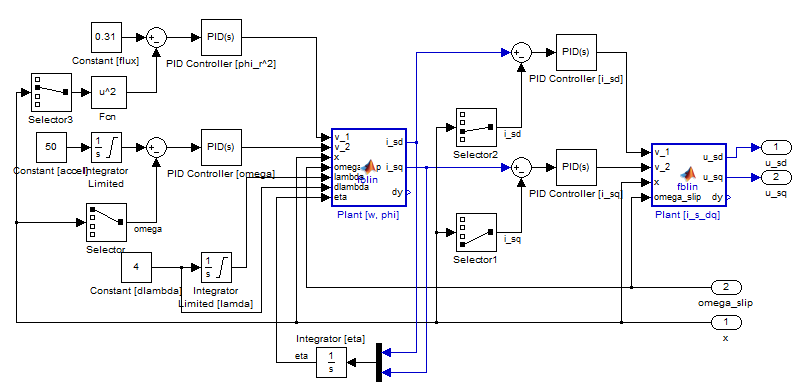}

Figure 6. Controller subsystem model.
\end{center}

To increase the adequacy of the modeling to the signals of the stator current $i_{sd}$ and $i_{sq}$ added additive Gaussian noise with variance 0.005 obtained from a source of pseudorandom numbers, which corresponds to the 15 mA random error of current measurement.

Parameters of the model motor are presented in Table 1.

\begin{tabular}{ | l | c | l | }
\hline 
   Symbol & Value & Description \\
   \hline
$P$ & 4 & rated power, kW \\
$M_{sr}$ & 0.175 & mutual inductance, H \\
$R_s$ & 1.2 & stator resistance, ohm \\
$R_r$ & 0.873 & rotor resistance, ohm \\
$L_s$ & 0.195 & stator inductance, H \\
$L_r$ & 0.195 & rotor inductance, H \\
$J$ & 0.013 & combined inertia, $kg \cdot m^2$ \\
$p$ & 2 & pole pairs \\
$T_m$ & 2 & load torque, N m \\
\hline 
\end{tabular}

Modeling results are shown on figures 7-9. From the data obtained it is clear that with the presence of noise in the current feedback control algorithm provides acceptable performance. In particular, the accuracy of speed control is 0,2 \% and accuracy of maintaining the magnetic flux is 2 \%.

\begin{center}
\includegraphics[width=0.8\textwidth]{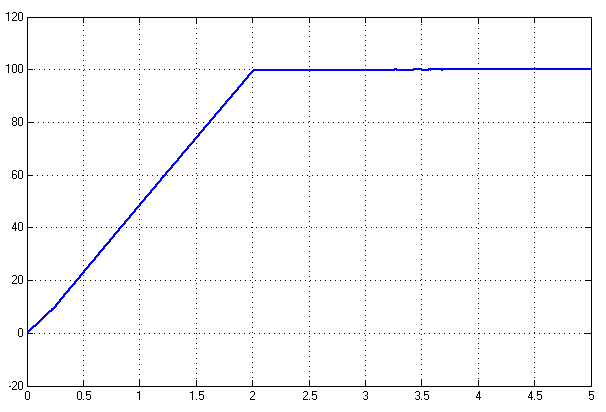}

Figure 7. Time plot of the motor speed $\omega(t)$. The objective of control is smooth acceleration to a speed 100 rad/sec within 2 seconds.
\end{center}

\begin{center}
\includegraphics[width=0.8\textwidth]{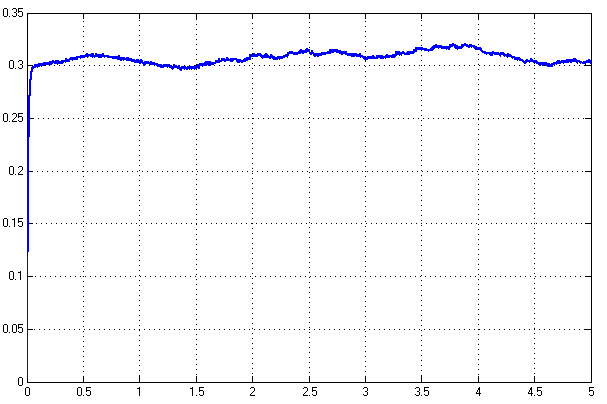}

Figure 8. Time plot of the rotor flux $\phi_r(t)$. The objective of control is maintain the magnetic flux of the rotor on the value of 0.31 Wb.
\end{center}

\begin{center}
\includegraphics[width=0.8\textwidth]{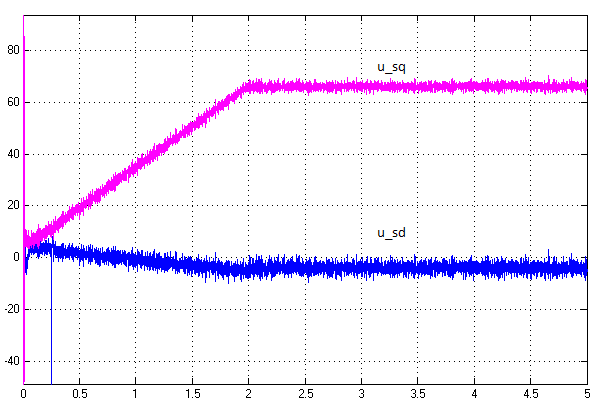}

Figure 9. Control inputs: the stator voltages $u_{sd}$ and $u_{sq}$.
\end{center}

\begin{center}
\includegraphics[width=0.8\textwidth]{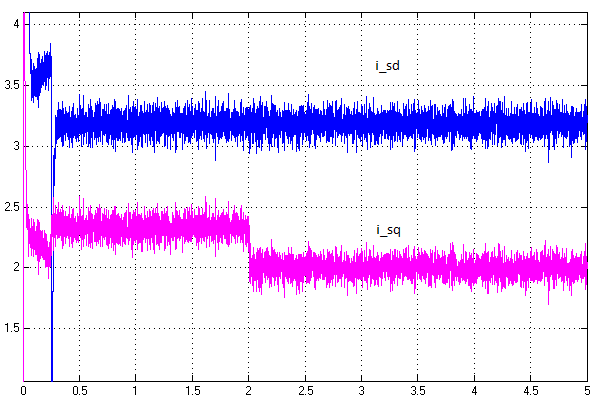}

Figure 10. Time plots of the stator currents $i_{sd}(t)$ and $i_{sq}(t)$.
\end{center}

\section{Three phase induction motor control: experimental implementation}

In this section we describe the results of experimental studies to verify and test the proposed approach drive control. General view of setup is shown in Figure 11. Three-phase asynchronous motor is mechanically connected to the controlled synchronous motor, which is used as a torque source. 

As a platform for implementing control algorithms used by the controller dSPACE DS5202, which is a hardware target for automatic code generation from MATLAB Simulink model. With technology of automatic code generation for hardware targe all implemented in Simulink algorithms were tested on a real induction motor without modifications. For the motor power supply power-stage based on frequency converter SEW MoviTrac is used. PWM control signals for transistors generated in dSPACE controller.

\begin{center}
\includegraphics[width=1.0\textwidth]{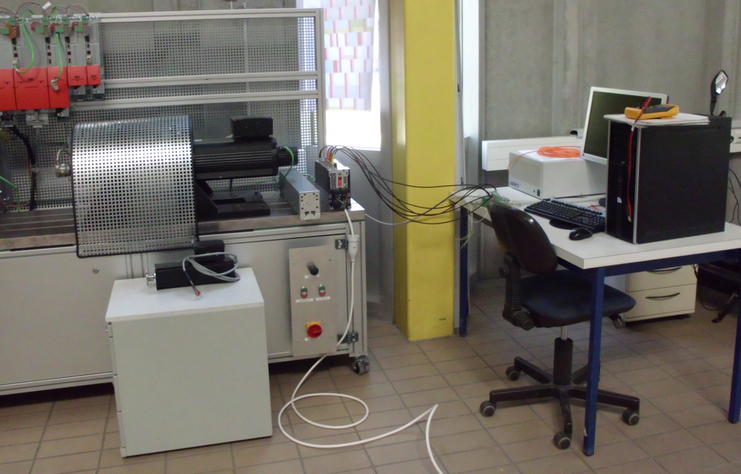}

Figure 11. Overall view of the experimental setup.
\end{center}

The experiment was replicated conditions similar to those used in the simulation, namely, the braking torque $T_m = 4.28$ N m. Tracking of speed setpoint $\omega_{ref}(t)$ was tested during experiments. The experimental results of speed control are shown in Figures 8-10. Torque was applied at moment $t = 32$ sec.

\begin{center}
\includegraphics[width=0.8\textwidth]{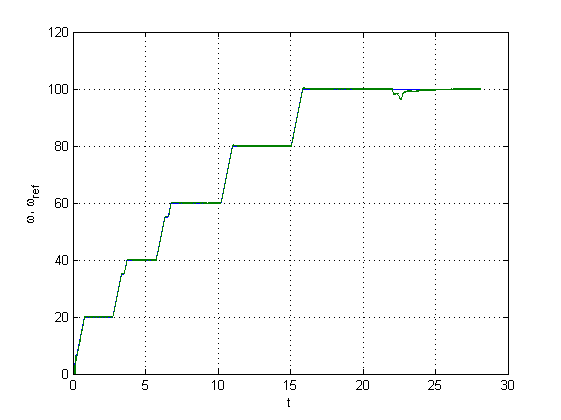}

Figure 12. Time plots of actual speed $\omega(t)$ and reference speed $\omega_{ref}(t)$.
\end{center}

\begin{center}
\includegraphics[width=0.8\textwidth]{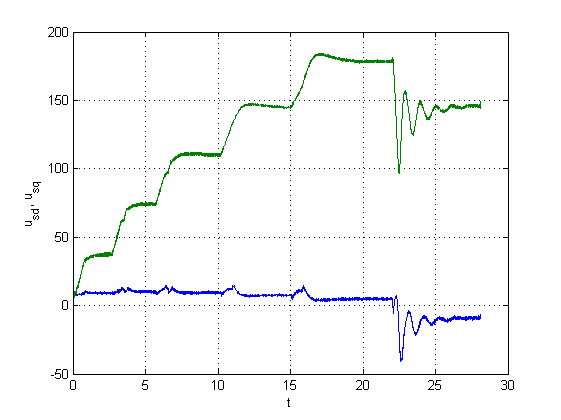}

Figure 13. Control inputs: the stator voltages $u_{sd}$ and $u_{sq}$.
\end{center}

\begin{center}
\includegraphics[width=0.8\textwidth]{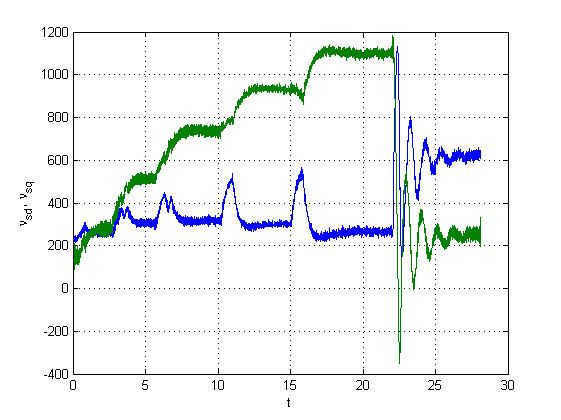}

Figure 14. Control outputs after the linearization $\nu_{sd}$ and $\nu_{sq}$.
\end{center}

\begin{center}
\includegraphics[width=0.8\textwidth]{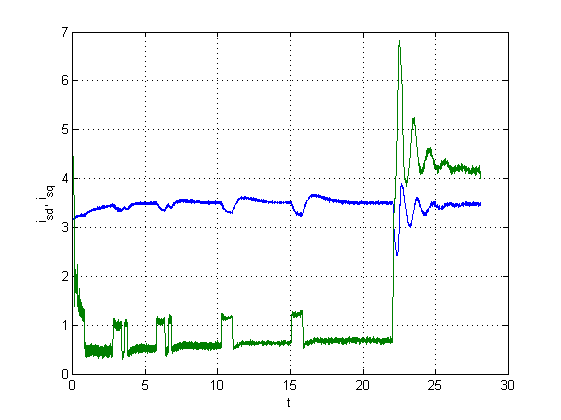}

Figure 14. Time plots of the stator currents $i_{sd}(t)$ and $i_{sq}(t)$.
\end{center}

From these results it is evident that the speed of rotation $\omega(t)$ is strictly corresponds to the set point $\omega_{ref}(t)$, the deviation does not exceed 0.8 \%.

\section{Conclusion}

In this paper we propose a new method for affine control systems, which combines the conceptual simplicity of feedback linearization methods and at the same time expands the scope of their applicability to irregular system with poorly expressed relative degree.

The method tested on an abstract system MIMO sistem with singularities in state space. Application of proposed method demonstrated on the speed and rotor magnetic flux control in the three-phase asynchronous motor. It has been modeled taking into account the effect of measurement errors of the stator currents. Also, the proposed approach is implemented and tested on an experimental setup.

Future work will focus on the investigation of uncertainties influence in an explicit form, the generalization of the approach using methods of differential geometry.

\end{document}